# How is Latin America engaging with responsible metrics? A systematic review comparing regional and global scientific production

Daniela Oyarzún-Cristi, Álvaro Cabezas-Clavijo, Alicia Moreno-Delgado

Universidad Internacional de la Rioja

Avenida de la Paz, 137, 26006, Logroño (Spain)

e-mail: doyarzunc@gmail.com

*This manuscript is a preprint and has not undergone peer review.*

**ABSTRACT**

Responsible metrics and responsible research assessment constitute an expanding field, driven primarily from the Global North through initiatives such as the San Francisco Declaration on Research Assessment (DORA), the Leiden Manifesto, and CoARA (Coalition for Advancing Research Assessment). However, despite growing international interest, Latin American scholarship on this subject remains largely unexamined. To address this gap, this paper presents a systematic review of the scientific literature on responsible metrics in the Latin American context. Drawing on 239 publications identified across five databases (Web of Science, Scopus, SciELO, Redalyc, and Dialnet) for the period 2012–2025, we compare the characteristics of Latin American output (n=80) with publications from the rest of the world included in the same corpus (n=159). The results show that Latin American output, concentrated in Brazil, Argentina, Mexico, and Colombia, has followed an irregular trajectory and lower volume than global production, with a peak between 2019 and 2022 followed by a sharp decline from 2023 onward, while output from the rest of the world has accelerated its growth during the same period. The Latin American literature is characterized by a predominance of theoretical and qualitative approaches, an orientation toward national and regional scales, and a strong alignment with the open science and open access agenda. In contrast, non-Latin American output displays greater methodological diversity and a growing number of applied studies documenting concrete reforms to research assessment systems. These asymmetries point to a gap between the normative debate, in which the region participates actively, and the empirical research oriented toward the operationalization of responsible assessment, in which its presence remains very limited. Overall, the findings point to the need to strengthen applied research on responsible metrics in Latin America and to better articulate local agendas with international debates on research assessment reform.

**Keywords:** Responsible metrics, Responsible research assessment, Research assessment reform, Latin America, DORA, CoARA, Systematic review.

## 1. Introduction

Research evaluation systems based on quantitative parameters have become central instruments for decision-making in science over the past few decades. Decisions about researcher hiring and promotion, funding allocation, institutional accreditation, and international rankings now depend to a considerable extent on bibliometric indicators derived from publications and citations (Aksnes et al., 2019). The appeal of this approach is straightforward: numerical proxies provide comparable and reproducible data at a fraction of the cost of qualitative review. Yet these evaluative models have been widely criticized for ignoring local realities and the structural inequalities present in diverse scientific contexts (Ràfols, 2019). They can also encourage strategic behavior, shifting attention away from research quality and toward the maximization of indicators, and fostering a conservative, risk-averse academic culture (de Rijcke et al., 2016).

In response to these limitations, international initiatives have emerged, driven primarily from Global North countries, aimed at reforming research evaluation practices. This movement, known as responsible metrics and defined as "a way of framing appropriate uses of quantitative indicators in the governance, management and assessment of research" (Wilsdon et al., 2015), has established itself as a new paradigm in research assessment. It sits within a broader conceptual framework, that of responsible research assessment (RRA), a term capturing different "approaches to assessment which incentivise, reflect and reward the plural characteristics of high-quality research, in support of diverse and inclusive research cultures" (Curry et al., 2020).

Despite growing international interest in responsible metrics and responsible research assessment, the Latin American context lacks systematic evidence that would allow for a comprehensive understanding of how this field has developed in the region. Gaps remain regarding its evolution, scope, characteristics, and relationship with global trends in evaluation reform. This paper addresses those gaps by mapping the responsible metrics movement in Latin America and characterizing the current state of research in the region. In doing so, it identifies areas of convergence and divergence with international debates, providing empirical evidence of relevance to research policy stakeholders, funding agencies, and higher education institutions seeking to advance more contextualized, inclusive, and responsible evaluation systems.

## 2. Background

### 2.1 Quantitative research evaluation and its limitations

Bibliometric indicators have been recognized as a fast, simple, and cost-effective decision-making tool, grounded in the apparent objectivity of numbers and supported by a quality standard widely accepted across the scientific community (Delgado-López-Cózar & Martín-Martín, 2019). These advantages rapidly led research evaluation systems worldwide to adopt citation-based metrics as proxies for quality and impact (Aksnes et al., 2019). Such metrics have since been applied across the entire research ecosystem: in the hiring, promotion, and tenure of researchers (Moher et al., 2018); in the evaluation of research and development proposals (Thamhain, 2014); in the assessment of research group performance within institutions (Coccia, 2005); in funding allocation (Abramo et al., 2009); and in the positioning of universities in global rankings (Johnes, 2018).

While quantification has proven attractive precisely because of its distance from subjective human judgment, the "tyranny of numbers" has fostered an obsession with control, accountability, and bureaucratization at the expense of research freedom (Bornmann & Marewski, 2024). New forms of governance built around rankings, quantitative assessments, bibliometrics, and impact measures have shaped the behavior of the "Homo academicus" (Bourdieu, 2008), leading researchers to align their activities with the criteria used in evaluation processes (Butler, 2003; Jiménez-Contreras et al., 2003). As a result, researchers are discouraged from pursuing risky or innovative work that demands more time or involves greater complexity (Alberts, 2013; Saura & Bolívar, 2019; Vera, 2017).

The pressure to publish in order to advance professionally has pushed researchers worldwide to embrace the well-known imperative of "publish or perish." This has contributed not only to academic burnout (Kulikowski et al., 2023), but also to the proliferation of questionable practices and metric manipulation, threatening academic integrity (Ma, 2022) and producing a contaminated scholarly record (Quaderi, 2023).

Peripheral scientific systems face an additional structural disadvantage. The major databases and indexing tools, designed and managed from the Global North, have built a model of academic prestige centered on journal impact factors and mainstream international rankings (Beigel & Salatino, 2015). As these criteria have been adopted globally, journals, languages, and scientific traditions that do not conform to that model have become systematically undervalued in research assessment processes (Salatino & Ruiz, 2021).

### 2.2 The emergence of responsible metrics

The concept of responsible metrics grew out of a series of initiatives calling for reform in research evaluation systems. From the San Francisco Declaration on Research Assessment (DORA, 2012), in which a group of scientists issued 18 recommendations addressed to funding agencies, academic institutions, journals, metric-related organizations, and researchers (Alberts, 2013), through to the Barcelona Declaration on Open Research Information (2024), a succession of voices has called for substantial changes in the way research is evaluated (Figure 1).

Two documents proved particularly influential. The Leiden Manifesto (2015), in which a team of bibliometrics experts formulated 10 principles for reforming research evaluation, argued that quantitative metrics should serve to support, rather than replace, expert qualitative judgment (Hicks et al., 2015). Around the same time, the report "The Metric Tide" (Wilsdon et al., 2015), commissioned by the British government, coined the term responsible metrics and proposed five defining dimensions: robustness, humility, transparency, diversity, and reflexivity.

Several European initiatives subsequently sought to operationalize these principles, including the European Commission's Next Generation Metrics report (2017) and the recommendations of the UK Forum for Responsible Research Metrics for the REF2021 evaluation exercise (Wilsdon et al., 2017). Another significant development was the creation in 2018 of the INORMS Research Evaluation Group, which later proposed the SCOPE Framework, a structured five-step model for conducting responsible research evaluations (Himanen et al., 2024). The 2019 Helsinki Initiative added a further dimension by promoting multilingualism in scholarly communication, calling for the recognition of publications in researchers' native languages and challenging the hegemony of English in research assessment (Federation of Finnish Learned Societies et al., 2019). The Hong Kong Principles (2020) extended this agenda by stressing the need to safeguard research integrity through the recognition of responsible practices, transparency in evaluation processes, open science, diversity of scholarly contributions, and a comprehensive view of research activity (Moher et al., 2020).

In 2022, the Coalition for Advancing Research Assessment (CoARA) renewed the push for systematic reform, advocating for fairer and more diverse evaluation systems that prioritize the responsible use of quantitative indicators and give central weight to qualitative peer judgment, recognizing the quality of diverse research outputs and practices (CoARA, 2022). This proposal generated debate and criticism, particularly regarding the potential displacement of traditional scientometrics (Abramo, 2024). In this context, some authors have argued for a balance between metrics, expert judgment, and science policy, rather than an outright replacement of quantitative approaches (Torres-Salinas et al., 2023). Tools such as narrative bibliometrics (Torres-Salinas et al., 2024) and narrative CVs (Bordignon et al., 2023; Cabezas Clavijo, 2024) have emerged as complementary approaches to traditional evaluation. Rather than eliminating quantitative indicators, they seek to moderate their relative weight, strengthen the contextualization of academic trajectories, and offer assessment frameworks more sensitive to different stages of the research career. Further global initiatives, including the UNESCO Recommendation on Open Science (2021) and the Barcelona Declaration on Open Research Information (2024), align with the responsible metrics agenda by promoting transparency, equity, and diversity in evaluation practices.

### 2.3 Responsible metrics in Latin America

The initial momentum toward responsible research assessment originated primarily in the Global North. However, prior to DORA, Latin America had already engaged in extensive debate about the use of mainstream scientific indicators applied to regional journals (López, 2013). While the region has made a notable and globally recognized contribution to the development of open access and open science (Millano-González, 2021), research evaluation policies have not followed the same trajectory.

The first regional milestone toward evaluation reform in Latin America came at the 2015 Assembly of the Latin American Council of Social Sciences (CLACSO). On that occasion, participants proposed rethinking evaluation from a critical and autonomous standpoint, promoting collaborative and solidarity-based practices as a means of addressing structural inequalities (CLACSO, 2015). In 2018, the Panama Declaration on Open Science sought to redefine assessment criteria by prioritizing scientific cooperation, knowledge networks, and capacity building, and by incorporating social and regional impact as a central dimension of evaluation (CILAC, 2018). That same year saw the emergence of AmeliCA, a collaborative Latin American initiative promoting and strengthening open science (Becerril-García, 2019), which encourages the use of responsible metrics and the adoption of CRIS systems for the evaluation of open science outputs (Vélez Cuartas et al., 2019).

In 2019, the Latin American Forum on Research Assessment (FOLEC) was established. It has since become a key actor at the global level, driving regional debate on the transformation of academic evaluation and seeking to orient research toward social needs from a more equitable, diverse, and inclusive perspective (Rovelli, 2022). The FOLEC Declaration of Principles (2022) reaffirmed the need to advance toward "a new academic and scientific evaluation for a socially relevant science in Latin America and the Caribbean," stressing the importance of adopting alternative approaches to research assessment in the region (CLACSO-FOLEC, 2022).

In this context, responsible evaluation has acquired a distinctly regional meaning in Latin America. Rather than simply adopting frameworks developed elsewhere, the challenge for the region is to build approaches that capture the full complexity of how knowledge is produced and circulated locally, encompassing open access, gender equity, teaching and outreach activities, societal engagement, and research integrity (Vommaro & Rovelli, 2022a). Yet despite the declarations and initiatives promoted at the regional level, and the formal commitments made by Latin American institutions as signatories of DORA, CoARA, and other international frameworks, no systematic body of evidence exists on how scholarly work on responsible metrics and responsible research assessment has developed among Latin American authors. This gap makes it difficult to assess the region's contribution to international debate, and to identify the challenges and opportunities involved in designing more relevant evaluation systems for Latin America.

## 3. Objectives

The general objective of this study is to identify and characterize the scientific output on responsible metrics in Latin American countries through a systematic literature review. To achieve this, six specific objectives (SO) are proposed:

SO1. To analyze the temporal evolution of scientific output on responsible metrics and responsible research assessment in Latin America.

SO2. To characterize the types of publications, countries of affiliation, institutional collaboration patterns, and leading authors involved in Latin American works.

SO3. To classify publications according to the stage of development at which contributions address topics related to responsible metrics and responsible research assessment in the Latin American context.

SO4. To identify the methodological approaches, levels of analysis, and target audiences addressed by the Latin American literature on responsible metrics.

SO5. To analyze the main thematic dimensions present in Latin American publications.

SO6. To compare publication patterns on responsible metrics in Latin America with production dynamics on this topic in the rest of the world.

**Figure 1. Timeline of milestones in responsible research assessment.**

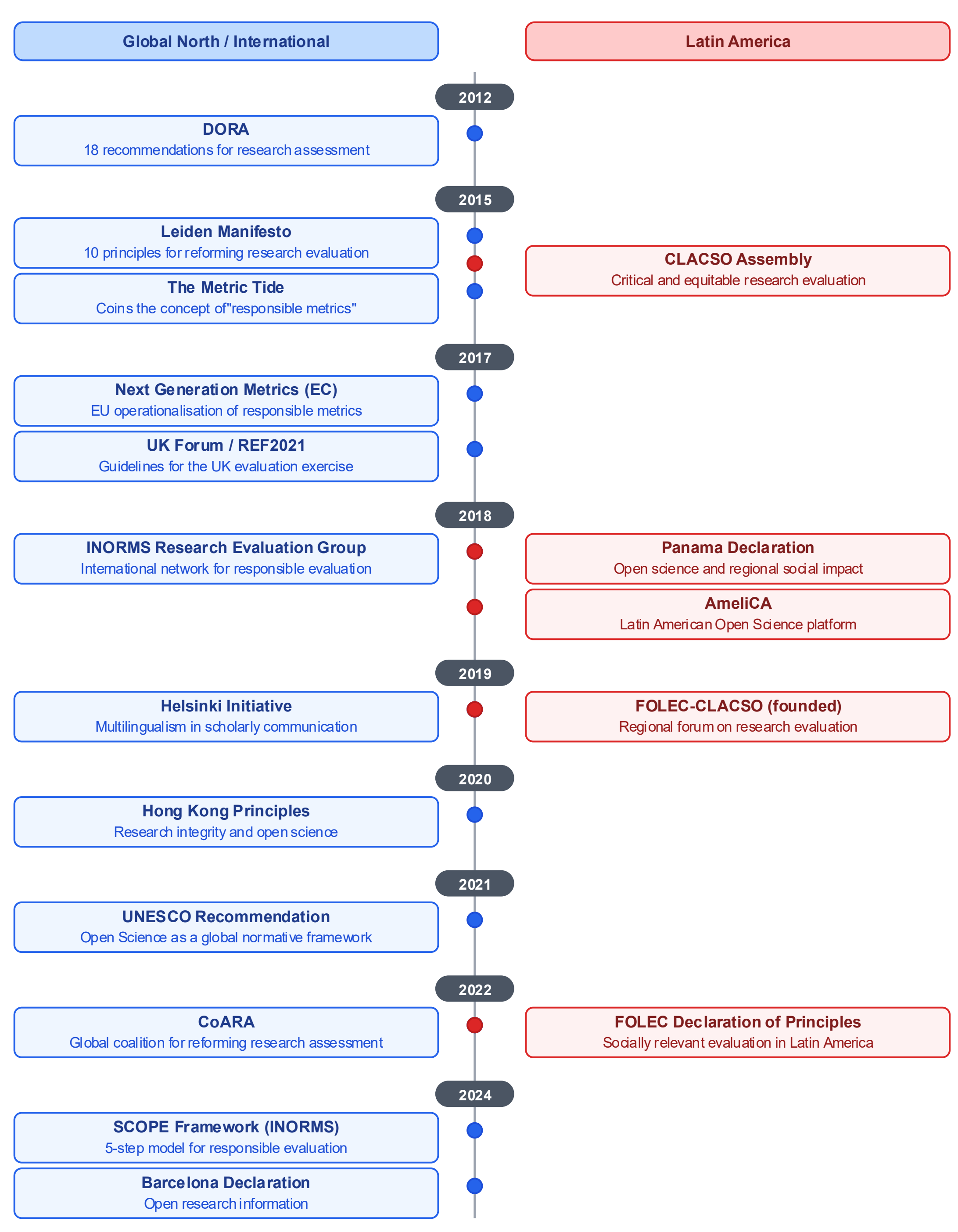

## 4. Methodology

A systematic literature review was conducted following the PRISMA 2020 guidelines (Preferred Reporting Items for Systematic Reviews and Meta-Analyses) (Page et al., 2021), in order to ensure transparency, reproducibility, and methodological rigor throughout the process of identifying, selecting, and analyzing the included studies. Given that the San Francisco Declaration on Research Assessment (DORA, 2012) represents a landmark in the field, the review covers the scientific literature published from that year onward. The analysis therefore spans the period 2012 to 2025.

### 4.1 Data sources and search strategy

Five databases were selected for the identification of scientific literature: Web of Science Core Collection (WoS), Scopus, SciELO, Redalyc, and Dialnet. The first two are multidisciplinary, internationally scoped databases widely used in bibliometric and research evaluation studies. SciELO, Redalyc, and Dialnet were included for their relevance to the visibility of Latin American output, particularly in Spanish and Portuguese.

Searches were carried out between March 8 and 26, 2026, using combinations of terms in English and Spanish related to responsible metrics and responsible research assessment. The search terms were: "Responsible metrics", "Responsible indicators", "Responsible research evaluation", "Responsible research assessment", "Responsible scientific evaluation", "Responsible scientific assessment", "Research Assessment Reform", "Research evaluation reform", "Narrative CV", "Declaration on Research Assessment", "The San Francisco Declaration", "Leiden Manifesto", and "Coalition for Advancing Research Assessment", connected using the Boolean operator OR.

### 4.2 Inclusion and exclusion criteria

To ensure broad coverage of the bibliographic record, all document types published in Spanish, English, or Portuguese were included. Records retrieved through the search that matched the search terms but had no relation to the subject of study were excluded, as were records lacking a title or abstract.

### 4.3 Study identification and screening process

Searches across the five selected databases yielded a total of 828 records, distributed as follows: Web of Science (n = 131), Scopus (n = 179), SciELO (n = 15), Dialnet (n = 87), and Redalyc (n = 416). All retrieved records were imported into the Zotero reference manager, where 175 duplicates were identified and removed, leaving 653 unique documents for initial screening. After applying the exclusion criteria, 386 records were assessed in full, and a further 147 were discarded for lacking thematic relevance. The final sample comprised 239 publications.

### 4.4 Geographic classification criteria

Since this study examines publications from both Latin America (LatAm) and the rest of the world (non-LatAm), the 239 records were classified manually according to the geographic origin of their authors, based on the institutional affiliation declared in each publication. Documents were assigned to the LatAm group if at least one author was affiliated with a Latin American institution; all remaining records were classified as non-LatAm. This process yielded 80 LatAm publications, of which 73 feature exclusively Latin American authorship, and 159 non-LatAm publications.

The complete identification, screening, eligibility, and inclusion process is summarized in the PRISMA 2020 flow diagram presented in Figure 2.

**Figure 2. Study identification, screening, eligibility, and inclusion process following PRISMA 2020.**

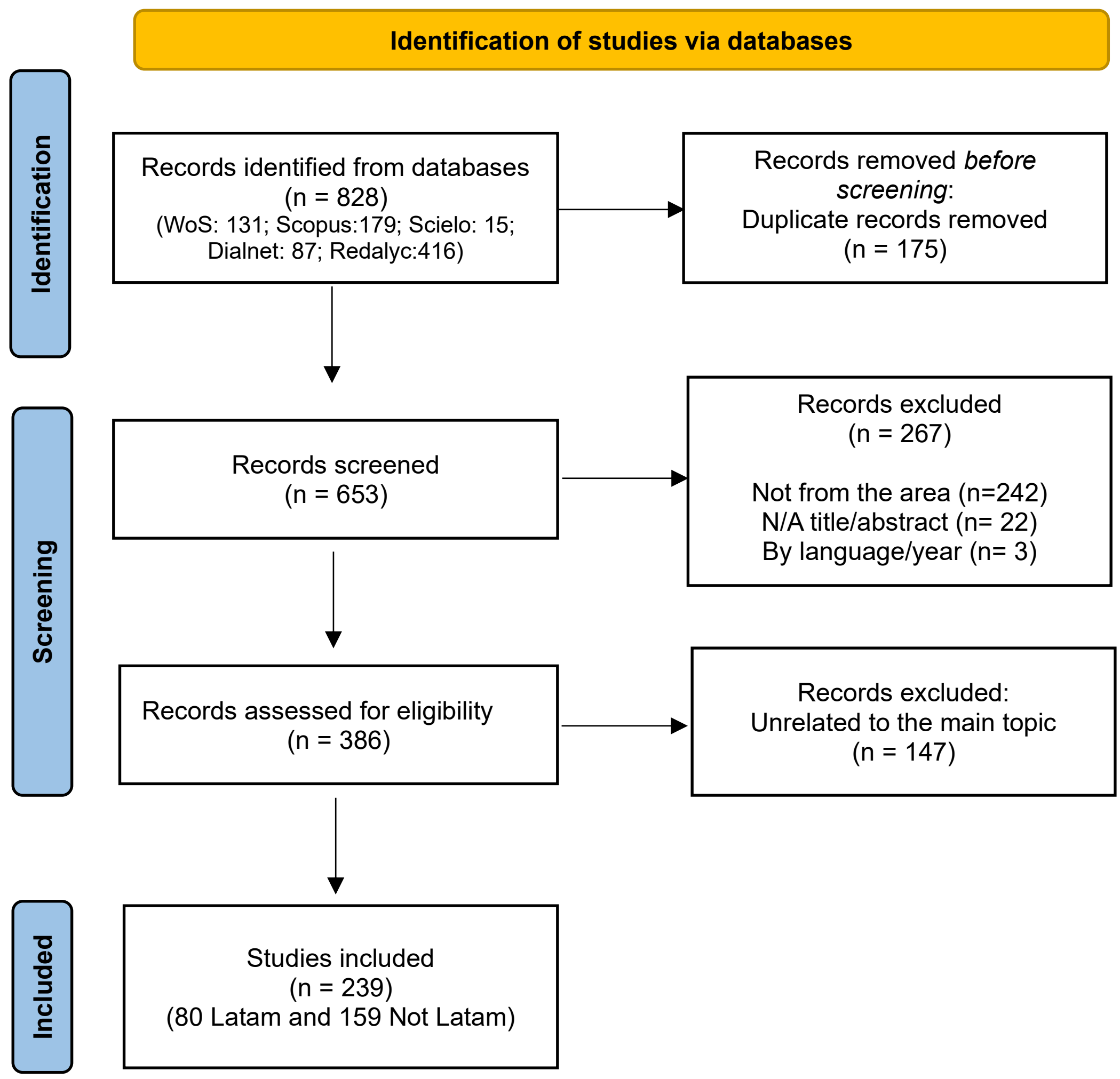

### 4.5 Descriptive analysis

The descriptive analysis covers the temporal evolution of publications, document types, countries of affiliation, institutional collaboration patterns, and the leading authors contributing to the study of responsible metrics.

For the temporal evolution and document type analyses, all 239 included records were used. For the remaining descriptive analyses, only documents classified as "Article" were considered (n = 139).

To characterize the geographic origin of contributions, the country corresponding to the first institutional affiliation declared by each author was used. In the count, each country was recorded only once per article, regardless of how many authors shared the same affiliation or how many institutions from the same country were involved.

For the analysis of institutional collaboration patterns, one affiliation per author was considered, selecting the first where more than one was listed. Only works involving two or more distinct affiliations were included, regardless of the number of contributors. Based on these data, a social network analysis was conducted using Gephi (version 0.11.2) (Bastian et al., 2009). Since institutional collaborations are bidirectional, the network was processed as an undirected graph. The ForceAtlas 2 layout algorithm was applied for visualization, and community detection was performed using the modularity algorithm of Blondel et al. (2008), which identifies clusters of nodes with greater internal than external connection density. Betweenness centrality was calculated using the Brandes (2001) algorithm, which measures how frequently a node appears on the shortest path between other nodes in the network, serving as an indicator of its role as a bridge or articulator between communities.

**4.6 Content analysis**

To assess the stage of development at which the responsible metrics topics addressed in the publications were situated, all 239 selected records were classified into three stages according to the criteria specified in Table 1.

**Table 1. Stages of development of publications related to responsible metrics and responsible research assessment identified in this study.**

| Stage | Criterios |
|---|---|
| 1. Critiques and calls for action | Publications that identify, examine, and/or draw conclusions about the limitations of quantitative evaluation and the journal impact factor, calling for improvements in assessment practices. These are primarily diagnostic in nature. |
| 2. Emergence of responsible metrics and responsible research assessment | Publications that explicitly use the terms "responsible metrics" and/or "responsible research assessment." Includes works proposing new models, metrics, and methodologies for reforming research evaluation. These are primarily theoretical in orientation. |
| 3. Responsible metrics in practice | Publications reporting the results of implementing reforms, metrics, or responsible evaluation systems in any area of research. These are primarily applied in nature. |

The remaining analyses, covering methodological approach, level of analysis, target audience, and thematic dimensions, were based on the 139 records classified as "Article." To characterize methodological approaches, this study draws on the classification proposed by Chu (2015) for research methodologies in the Social Sciences. All methodologies identified in each article were included in the analysis, even where more than one was present in a single work. Table 2 presents the analytical dimensions, including methodology, level of analysis, and target audience, along with the possible categories into which each work may be classified and their corresponding descriptions.

**Table 2. Dimensions, categories, and classification criteria for works related to the study of responsible metrics.**

| Dimension | Categories | Criterium |
|---|---|---|
| ***Research method*** | Theorical approach | Literature reviews, scoping reviews. |
| | Bibliometrics | Studies based on bibliometric indicators, citation analysis, webometrics, and altmetrics. |
| | Content analysis | Studies based on documentary and textual analysis. |
| | Interview | Individual and group interviews with stakeholders. |
| | Experiment | Hypothesis testing using a study group. |
| | Questionnaire | Surveys administered to stakeholder groups. |
| | Other | Case studies, meta-analyses, proposals for new methodologies, observation, data analysis, inductive and deductive analysis, think-aloud protocol, analytical approach. |
| ***Level of analysis*** | Global | General studies on evaluation systems. |
| | Regional | Studies focused on Latin America or other regions. |
| | Country | Studies focused on a specific country. |
| | Institution | Studies on universities, research centers, or libraries. |
| ***Target audience*** | Policy makers | Research policy stakeholders. |
| | Funders | Funding agencies. |
| | Publishers | Scientific journal publishers and editors. |
| | Institutions | Universities, research centers, and libraries. |
| | Researchers | Researchers and academics. |
| | Evaluators | Actors involved in research assessment. |

To quantify and compare the diversity of distributions across categories in dimensions with mutually exclusive classifications, the Shannon-Wiener diversity index (H) was calculated for Level of Analysis and Target Audience. This index measures how evenly observations are distributed across categories, ranging from $H = 0$ (total concentration in a single category) to $H_{max} = \ln(k)$, where k is the number of categories. To test whether the differences observed between LatAm and non-LatAm publications were statistically significant, a Pearson chi-square test was applied to the same two dimensions, with a significance threshold of $p < 0.05$. Research Method was excluded from both analyses, as articles could be assigned to more than one methodological category simultaneously, which invalidates both the Shannon index and the chi-square test. All chi-square analyses were performed using GraphPad Prism (version 8.4.2). The Shannon-Wiener index was calculated in Microsoft Excel. Given its descriptive nature, it does not depend on sample size for its validity; however, the values reported reflect the distribution observed in the analyzed corpus and should not be interpreted as population-level estimates, particularly given the relatively small size of the Latin American sample ($n = 41$).

A chi-square test of independence was also applied to assess whether the distribution of publications across development stages (Stages 1, 2, and 3; see Table 1) differed significantly between LatAm and non-LatAm, using the full set of 239 records included in the study. To complement this analysis, effect size was calculated using Cramér's V, and standardized residuals were examined for each cell to identify the region-stage combinations contributing most to the observed association, with $p < 0.05$ considered statistically significant.

Finally, the thematic classification of publications on responsible metrics and responsible research assessment yielded a total of 30 relevant topics. To facilitate their systematization, comparative analysis, and presentation of results, these topics were grouped into six analytical dimensions (Table 3).

**Table 3. Thematic dimensions identified in articles related to responsible research assessment.**

| Nº | Dimension | Criteria | Topics |
|---|---|---|---|
| 1 | Scholarly Production & Open Research Communication | Generation and dissemination of knowledge | Scientific publications; Research output; Digital repositories; Open science; Open access; Data management. |
| 2 | Traditional Evaluation Metrics | Classical quantitative indicators | Bibliometrics; Scientometrics; Impact factor; Citations; Quantitative evaluation. |
| 3 | Alternative and Emerging Metrics | New ways of measuring impact | Altmetrics; Social media; New metrics; Social impact of science; SDGs. |
| 4 | Research Evaluation Systems, Frameworks & Policies | Evaluation structures and governance | Evaluation frameworks; Evaluation reform; Responsible assessment; Evaluation criteria/indicators; University rankings. |
| 5 | Qualitative & Career-Based Assessment | Quality- and narrative-centered evaluation | Qualitative indicators; Narrative CV; Narrative bibliometrics; Academic/research career; Research quality. |
| 6 | Integrity, Ethics, Equity & Inclusion | Ethical, social, and technological dimensions | Retractions; Research ethics; AI; Gender in research. |

## 5. Results

### 5.1 Temporal evolution

This systematic literature review identified a total of 239 publications related to the study of responsible metrics between 2012 and 2025. Of these, 80 are affiliated with Latin American institutions (33.5%), while 159 originate from the rest of the world (66.5%) (Figure 3).

Latin America joined the debate on research evaluation reform from 2013 onward, though its contributions have not been sustained over time. LatAm output reached its most productive period between 2019 and 2022, with 12 works published in each of 2019, 2021, and 2022 (representing 45% of the 80 LatAm works. From 2023 onward, however, output has fallen steadily, with between 4 and 6 works per year through 2025, the lowest levels recorded across the entire series alongside 2015.

Non-LatAm publications, by contrast, show a growing trend from 2018 onward that accelerates markedly in the final three years: from 14 works in 2023 to 24 in 2024 and 32 in 2025, the latter being the highest annual figure recorded for either group across the entire period.

The combined trajectory reflects an overall upward trend driven increasingly by non-LatAm output. Whereas both groups contributed in roughly equal measure to the total between 2019 and 2022, growth since 2023 is explained almost entirely by the rise in non-LatAm publications. Two divergent patterns thus emerge: while research interest in the topic has declined in Latin America after 2022, it has intensified rapidly elsewhere.

**Figure 3. Temporal evolution of publications on responsible metrics in LatAm and non-LatAm contexts, 2012–2025.**

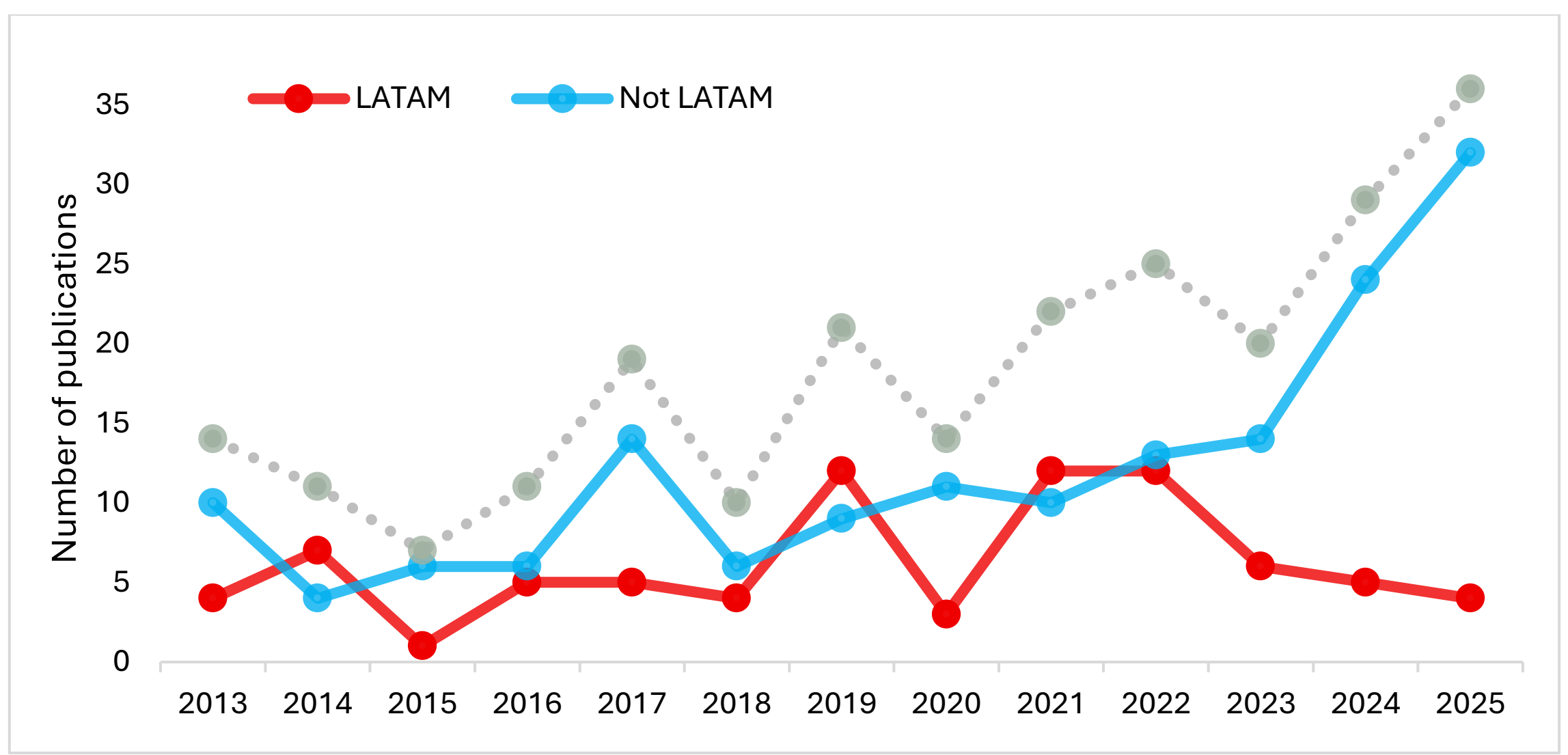

### 5.2 Publication types

In both contexts, journal articles are the predominant document type, accounting for 51.2% of LatAm-affiliated documents (41 publications) and 61.6% of non-LatAm documents (98 publications). Among the document types identified, a notable feature is the high proportion of contributions expressing authors' opinions (editorials, comments, letters, and opinion pieces), which do not constitute original research outputs: these account for 36.2% of LatAm publications (29 documents) compared to 25.8% in the non-LatAm group (41 documents). The remaining document types (books, chapters, communications, conference papers, dossiers, essays, reports, seminars, and workshops) represent a similar share in both contexts (12.5% LatAm; 12.6% non-LatAm). A chi-square test of independence applied to these three grouped categories (Articles;

Editorials/Comments/Letters/Opinion pieces; and Other) found no statistically significant differences between LatAm and non-LatAm in the distribution of publication types ($\chi^2$ = 2.98; df = 2; p = 0.226; Cramér's V = 0.11). This indicates that both contexts share a broadly similar document profile, despite the higher relative proportion of articles in the non-LatAm group and of opinion-type documents in the LatAm group.

### 5.3 Country of affiliation

To determine the countries of affiliation with the highest number of articles, the 139 identified articles were considered. Within Latin America, Brazil recorded the largest number of articles (13) and stands out for the consistency of its output over time. Argentina follows with 10 articles, though its most recent publication dates from 2023, while Mexico and Colombia, with 8 works each, show an irregular publication pattern. Outside Latin America, Spain leads with the highest number of publications (21), followed by Germany (14), Canada (11), the Netherlands (10), the United Kingdom (10), and Italy (9).

### 5.4 Leading authors

Of the 327 authors identified, 105 are affiliated with Latin American institutions (32%), while the remaining 222 hold other affiliations (68%). Two authors account for the highest number of articles (4) in the field of responsible metrics: Gabriel Vélez-Cuartas from the Universidad de Antioquia (Colombia), and Alexander Rushforth from Leiden University (the Netherlands). A further 11 authors contributed three works each, representing 3.4% of the total (Table 4). Additionally, 37 authors have two publications (11.3%), while the remaining 277 authors contributed a single work each, accounting for 84.7% of the total. In terms of publication output per author, LatAm and non-LatAm groups show similar distributions, and no significant differences were found between them.

**Table 4. Authors with the highest number of published works on responsible research assessment, 2012–2025.**

| Author | Affiliation | Country | N publications |
|---|---|---|---|
| Rushforth, Alexander | Leiden University | Netherlands | 4 |
| Vélez-Cuartas, Gabriel* | Universidad de Antioquia | Colombia | 4 |
| Aguado-López, Eduardo* | Universidad Autónoma del Estado de México | Mexico | 3 |
| Chinchilla-Rodríguez, Zaida | Consejo Superior de Investigaciones Científicas | Spain | 3 |
| Daraio, Cinzia | University of Ottawa | Italy | 3 |
| Himanen, Laura | Tampere University | Finland | 3 |
| Kullmann, Sylvia | DIPF Leibniz | Germany | 3 |
| Rousseau, Ronald | University of Antwerp | Belgium | 3 |
| Sivertsen, Gunnar | Nordic Institute for Studies in Innovation, Research and Education | Norway | 3 |
| Suárez Tamayo, Marcela* | Universidad de Antioquia | Colombia | 3 |
| Torres-Salinas, Daniel | Universidad de Granada | Spain | 3 |
| Uribe-Tirado, Alejandro* | Universidad de Antioquia | Colombia | 3 |
| Zhang, Lin | The North China University of Water Resources and Electric Power | China | 3 |

*Authors with LatAm affiliation.*

### 5.5 Institutional collaboration

Of the 41 Latin American articles, 46% involved inter-institutional collaboration (19 articles), while among the 98 non-Latin American articles this proportion reached 41% (40 articles). Based on the 59 publications with inter-institutional collaboration, a network analysis was conducted using Gephi (version 0.11.2). In an undirected graph, 133 nodes (collaborating institutions) and 215 edges (connections between institutions) were identified. The modularity analysis (Blondel et al., 2008) detected 37 communities (modularity = 0.852) with a marked geographic pattern, revealing that collaboration occurs primarily within national boundaries (Figure 4).

**Figure 4. Network analysis of institutional collaborations.**

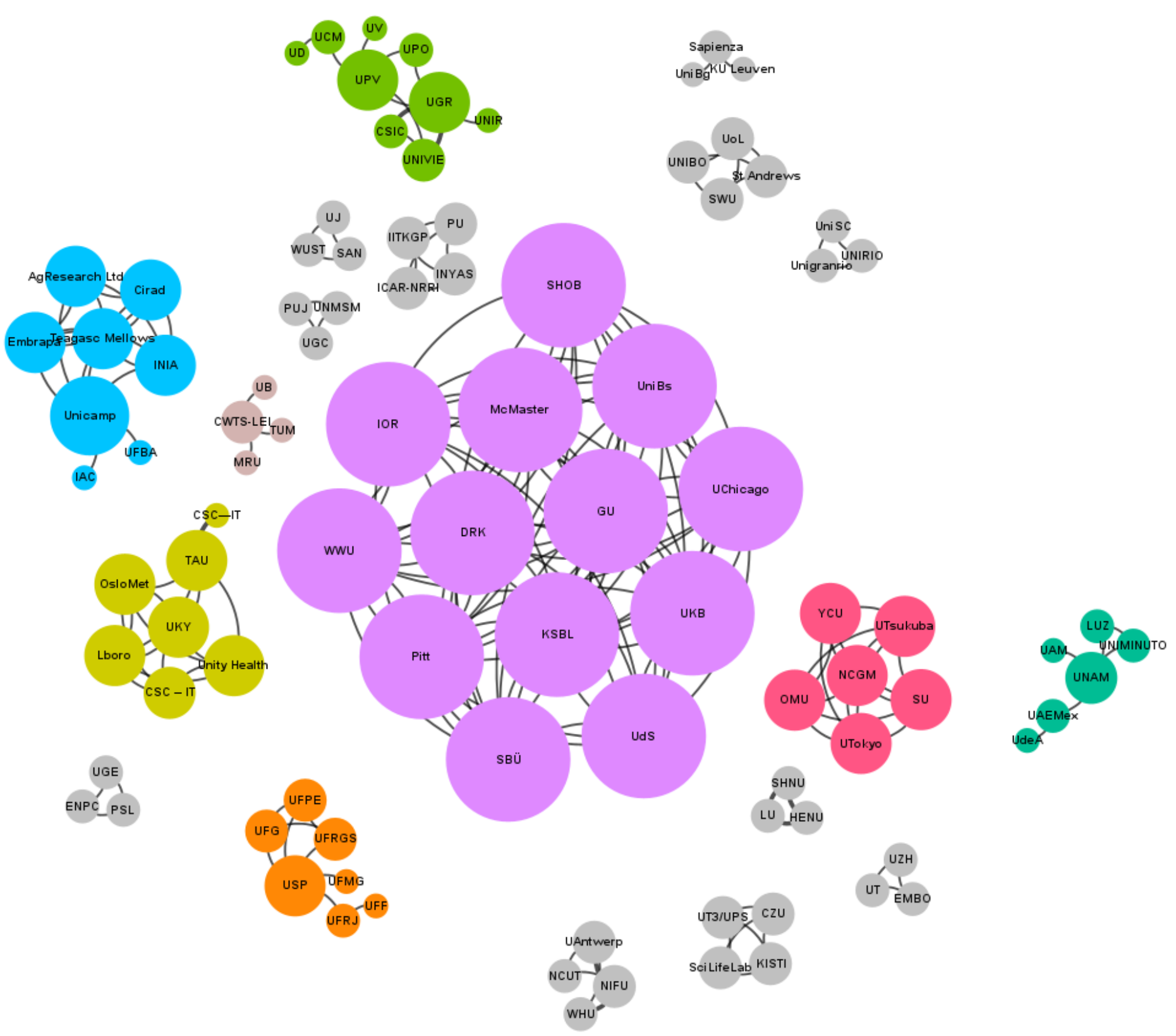

The purple community, the largest and most central, groups European and North American institutions with high output in responsible assessment. The green cluster brings together Spanish and Central European institutions (UPV, UGR, CSIC), while the blue cluster integrates international centers alongside Brazilian universities (Unicamp, UFBA). Two communities with a strong Latin American presence also stand out: an orange cluster of Brazilian universities (USP, UFG, UFRGS, UFMG, UFRJ, UFF) and a turquoise cluster comprising Mexican, Colombian, and Venezuelan institutions (UNAM, UAEMex, UdeA, LUZ), both operating relatively independently within the global network. Betweenness centrality analysis shows that the Spanish UPV is the primary bridging node in the network, followed by the Brazilian institutions Unicamp and USP. This is a noteworthy finding, as it suggests that a significant share of collaborations linked to Latin America are articulated through European nodes, even though consolidated Latin American national networks (Brazil, Mexico-Colombia) also exist and operate with a considerable degree of independence.

### 5.6 Stage of development in the discussion of responsible metrics

To examine how contributions to responsible metrics have evolved, all publications were classified into three stages: Stage 1 (Critiques and calls for action), Stage 2 (Emergence of responsible metrics and RRA), and Stage 3 (Responsible metrics in practice).

**Figure 5. Temporal evolution of publications on responsible metrics in LatAm (a) and non-LatAm (b) contexts, by stage of development**

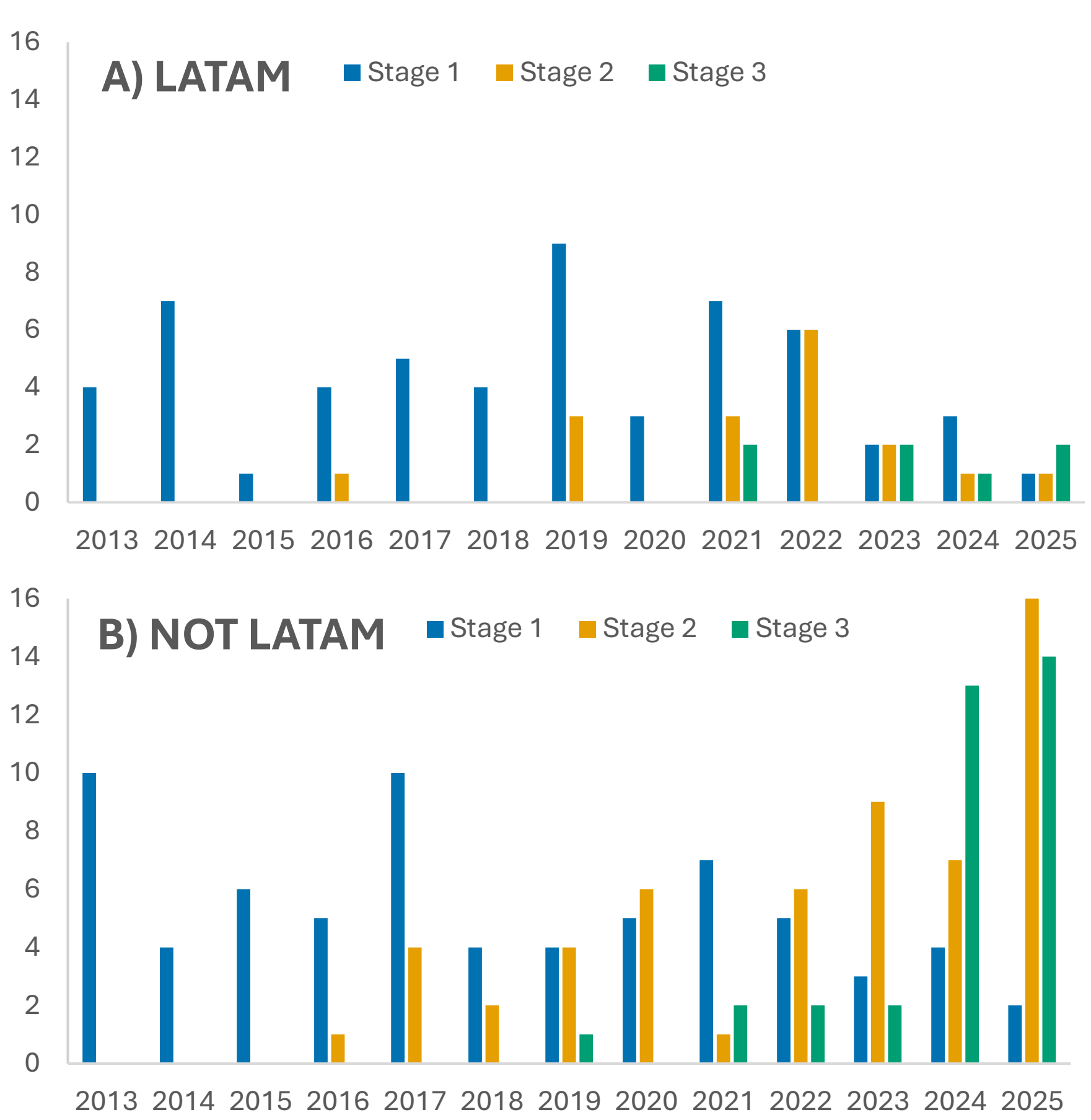

This classification yielded 125 works assigned to Stage 1 (52.3%), 73 to Stage 2 (30.5%), and 41 to Stage 3 (17.2%). Breaking down by context, LatAm output is heavily concentrated in Stage 1, which accounts for 70.0% of publications (56 works), followed by Stage 2 (21.2%; 17 works) and Stage 3 (8.8%; 7 works). The non-LatAm distribution is more balanced: Stage 1 (43.4%; 69 works), Stage 2 (35.2%; 56 works), and Stage 3 (21.4%; 34 works). Taken together, Latin American literature remains concentrated in the questioning and diagnostic phase (Stage 1), while non-LatAm output shows a more advanced transition toward the development of proposals (Stage 2) and, above all, their practical application (Stage 3). This difference is statistically significant ($\chi^2 = 15.56$; df = 2; $p < 0.001$; Cramér's V = 0.26), with the overrepresentation of LatAm in Stage 1 being the primary driver of the result (residual = 2.19; $p < 0.05$).

### 5.7 Research method, level of analysis, and target audience

The content analysis revealed further differences between LatAm and non-LatAm contexts (Table 5). Theoretical approaches constitute the predominant method in LatAm, accounting for 73% of articles, compared to 55% in the rest of the world. Bibliometric studies rank second at 17%, a proportion higher than in non-LatAm publications (10%). Content analysis and interviews (7% each in LatAm) are more common in non-LatAm research (12% and 10% respectively). Overall, non-LatAm articles draw on a broader range of methods, including case studies and works proposing new methodologies and evaluation frameworks.

Regarding level of analysis, LatAm publications are oriented primarily toward national and regional studies (39% and 31.7% respectively), while non-LatAm articles concentrate at the global level (50.0%). The Shannon-Wiener index confirms that LatAm shows a more balanced distribution across levels of analysis (H = 1.290; 93% of H_max = 1.386) than non-LatAm (H = 1.139; 82%), where concentration at the global level reduces overall diversity. The chi-square test confirms that these differences are statistically significant ($\chi^2 = 15.68$; df = 3; $p = 0.001$).

**Table 5. Research method, level of analysis, and target audience in articles related to responsible research assessment (2012–2025).**

| | | LATAM | | NOT LATAM | | Total | |
|---|---|---|---|---|---|---|---|
| **Dimension** | **Categories** | **n** | **%** | **n** | **%** | **n** | **%** |
| Research method | Theorical approach | 30 | 73% | 54 | 55% | 84 | 53% |
| | Bibliometrics | 7 | 17% | 10 | 10% | 17 | 11% |
| | Content analysis | 3 | 7% | 12 | 12% | 15 | 9% |
| | Interview | 3 | 7% | 10 | 10% | 13 | 8% |
| | Experiment | 2 | 5% | 1 | 1% | 3 | 2% |
| | Questionnaire | 1 | 2% | 6 | 6% | 7 | 4% |
| | Other | 1 | 2% | 18 | 18% | 19 | 12% |
| Level of analysis | Global | 7 | 17% | 49 | 50% | 56 | 40% |
| | Regional | 13 | 32% | 12 | 12% | 25 | 18% |
| | Country | 16 | 39% | 31 | 32% | 47 | 34% |
| | Institution | 5 | 12% | 6 | 6% | 11 | 8% |
| Target audience | Policy makers | 13 | 32% | 27 | 28% | 40 | 29% |
| | Funders | 2 | 5% | 3 | 3% | 5 | 4% |
| | Publishers | 7 | 17% | 15 | 15% | 22 | 16% |
| | Institutions | 7 | 17% | 26 | 27% | 33 | 24% |
| | Researchers | 4 | 10% | 16 | 16% | 20 | 14% |
| | Evaluators | 8 | 20% | 11 | 11% | 19 | 14% |

*Note: percentages for the Research Method dimension do not sum to 100%, as each article could be assigned to more than one methodological category simultaneously.*

As for target audience, the results show partially convergent patterns between both contexts. In LatAm, publications are directed primarily at research policy makers, followed by publishers, institutions, and evaluators. In non-LatAm, policy makers are also the primary target audience, though institutions and researchers carry greater relative weight than in LatAm. Both the Shannon index (H LatAm = 1.661; H non-LatAm = 1.643; H_max = 1.792) and the chi-square test ($\chi^2$ = 3.905; df = 5; p = 0.563) indicate no statistically significant differences between the two contexts in terms of target audience distribution. This suggests that both address a similar range of audiences, and that the differences observed reflect variations in the relative weight of each group rather than differential concentration.

### 5.8 Main thematic dimensions

Latin American attention has focused primarily on "Scholarly Production & Open Research Communication" (39.0%), followed by "Research Evaluation Systems, Frameworks & Policies" (24.4%). In the non-LatAm context, the pattern is reversed: articles lean more toward evaluation frameworks and policies (33.7%), with topics related to scholarly production and communication ranking second (19.4%). The dimensions covering "Traditional Evaluation Metrics," "Alternative and Emerging Metrics," and "Qualitative & Career-Based Assessment" show comparably modest interest across both contexts, while "Integrity, Ethics, Equity & Inclusion" accounts for the smallest number of articles in both groups (Figure 6).

**Figure 6. Distribution of thematic dimensions identified in articles published between 2012 and 2025.**

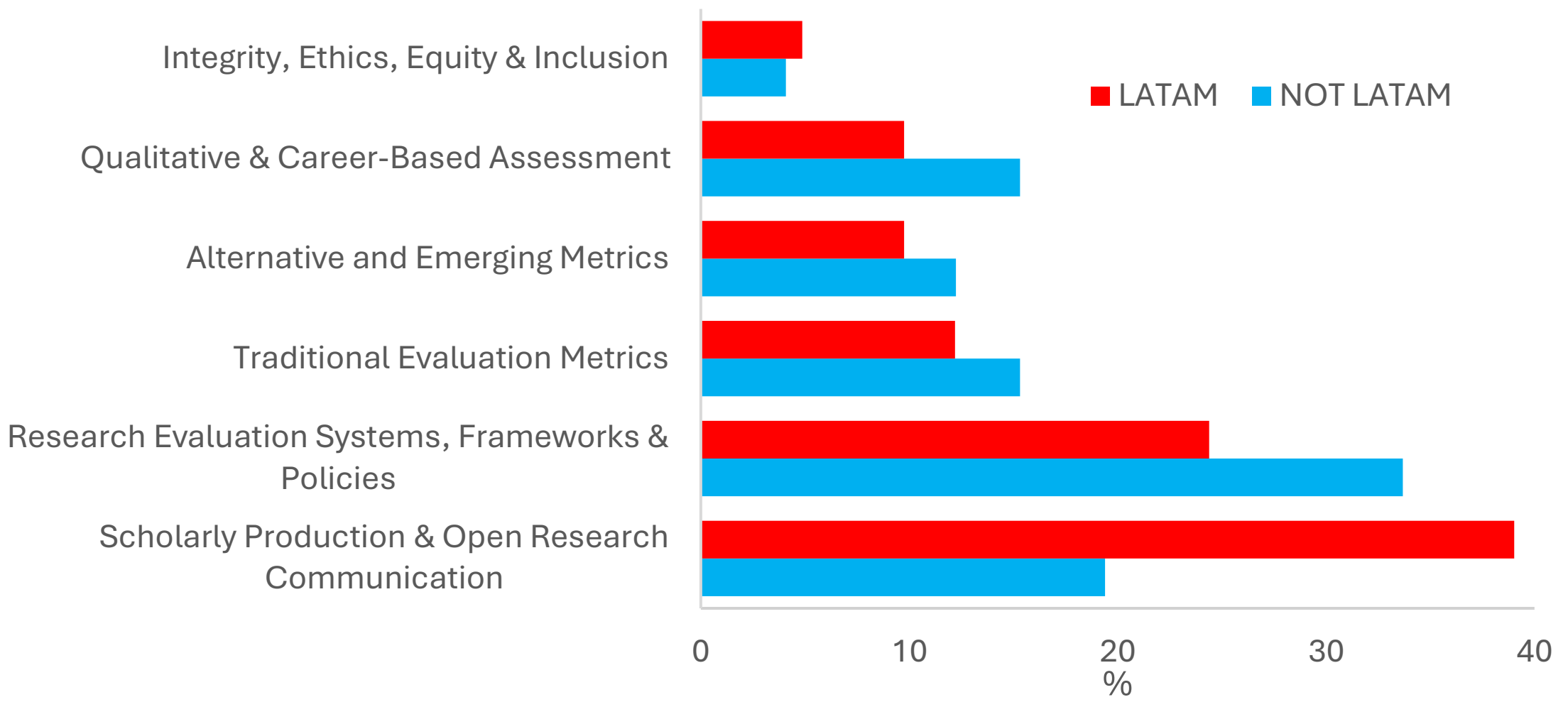

*Note: Values on the horizontal axis are expressed as percentages calculated over the total number of articles in each group (LatAm: n=41; non-LatAm: n=98). A single article may be associated with more than one thematic dimension.*

This is also worth noting: of the 30 topics identified, only half (16) appear in Latin American research, whereas non-LatAm publications cover 28 (93% of all identified topics). Topics such as narrative CVs, narrative bibliometrics, SDGs, social media, repositories, and the social impact of science are absent from Latin American articles.

## 6. Discussion and conclusions

The analysis of Latin American scientific output on responsible metrics reveals a progressive but uneven incorporation of the region into this field. While the topics addressed show points of convergence with international debate, significant asymmetries persist in key dimensions, including the pace of temporal growth, publication types, collaboration networks, methodological approaches, levels of analysis, and thematic coverage.

Addressing the first specific objective (SO1), Latin American publications on responsible metrics have followed an irregular, intermittent trajectory with lower volume than global output. This pattern may be linked to the historically limited presence of regional initiatives aimed at reforming research evaluation. As of 2021, only 15 global initiatives on responsible assessment had been identified, of which just one represented Latin America: FOLEC (Daraio & Vaccari, 2022). While CLACSO-FOLEC has established itself since its creation in 2019 as a relevant actor in regional debate, aligned with Open Science, the 2030 Agenda, and good evaluation practices (Rovelli, 2022), the data show that regional output has declined sharply from 2023 onward. This stands in contrast to the sustained growth observed elsewhere, which reached its highest values precisely in 2024 and 2025.

Regarding the characterization of publications, countries of affiliation, authors, and collaboration patterns (SO2), the Latin American literature is distinguished by a high proportion of reflective and argumentative texts, such as editorials, comments, and letters (36% of regional output), oriented toward problematizing the adoption of responsible metrics and the reform of university evaluation systems (Morales Perdomo, 2022; Vélez Cuartas, 2022; Vommaro & Rovelli, 2022a). These works address responsible assessment in relation to open science (Melero et al., 2023), ethical data management (Belmar & Atenas, 2024), and editorial alignment with DORA (Arias, 2021). Outside Latin America, while reflective publications are less frequent, they address similar issues from specific national or institutional perspectives (Delgado López-Cózar et al., 2021; Dirnagl & Pariente, 2024; Rafols & Molas Gallart, 2022).

From a geographic perspective, Brazil accounts for the largest share of regional output, with a focus on responsible assessment and equity (da Silva et al., 2024, 2025; Kowaltowski et al., 2021). Argentine contributions concentrate on open science and evaluation systems for researchers (Atrio, 2018; Batthyány et al., 2023; Beigel, 2022; Salatino & Macedo, 2022; Vommaro & Rovelli, 2022b), Mexican scholarship tends to address publication circuits and journals, as well as academic career evaluation (Aguado-López & Becerril-García, 2021; Arévalo-Guízar & Rueda-Beltrán, 2016; Vasen & Vilchis, 2017), while Colombia has developed contextualized proposals for responsible assessment, linked to national open science policies and alternative models for evaluating scientific journals (Uribe et al., 2023; Vélez Cuartas et al., 2019, 2021).

Authorship in this field is predominantly circumstantial: only 13 authors have published at least three works during the period, with no notable differences between the two groups. Institutional collaboration patterns among Latin American researchers are weak and fragmented, with a tendency toward single-institution co-authorship.

Regarding the stage of development of contributions (SO3), publications addressing critiques and calls for action remain present across all contexts (Aguillo, 2021; Aranda, 2022; Delgado López-Cózar & Martín Martín, 2019; Hatch & Curry, 2020; Jacobs, 2024). However, significant differences emerge between the two groups. Some 70.0% of Latin American publications fall within Stage 1 (56 works), compared to 43.4% from the rest of the world (69 works). Only 7 Latin American works (8.8% of regional output) correspond to Stage 3, which captures applied experiences of responsible assessment, against 34 from the rest of the world (21.4%). This asymmetry shows that Latin America remains largely in a normative debate phase, while other regions have begun to operationalize and document concrete reforms. These findings point to the need for the Latin American research community to direct greater effort toward empirical studies capable of guiding the various actors involved in science management and evaluation, and of fostering the adoption of new assessment practices in regional institutions, along the lines of developments seen in the Netherlands (Rushforth, 2025) or China (Shu et al., 2022; Zhao et al., 2025).

Regarding methodological approaches, levels of analysis, and target audiences (SO4), Latin American output shows a clear preference for theoretical and qualitative approaches, with a predominance of conceptual and discursive analysis. Non-LatAm literature, by contrast, displays greater methodological diversity, incorporating empirical studies, institutional diagnostics, and practical applications aimed at assessing the perceptions and experiences of academics and researchers regarding evaluation systems (Kulikowski et al., 2023; Robinson-Garcia et al., 2023).

Latin American literature is also characterized by studies of national and regional scope, directed primarily at decision-makers and public policy stakeholders. This orientation reflects the efforts of the regional academic community to influence the transformation of research evaluation systems (Arellano-Rojas et al., 2022; Calisto-Breiding et al., 2021; Vommaro & Rovelli, 2022a). In the non-LatAm context, global and national analyses predominate, addressing researchers and academics on topics such as academic careers, wellbeing, narrative CVs, and perceptions of evaluation (Bhakta et al., 2025; de Herde et al., 2021; Saura & Bolívar, 2019).

With respect to thematic dimensions (SO5), studies on evaluation systems and policies and on metrics and impact predominate in both Latin America and internationally, in line with debates surrounding DORA (Orduña-Malea & Bautista-Puig, 2024). However, Latin American literature shows a particularly strong orientation toward scholarly communication, open science, and open access, shaped by both regional tradition and recent international recommendations (UNESCO, 2021; Budapest Open Access Initiative, 2022). In the non-LatAm context, works on reference frameworks and operational models for responsible assessment stand out, such as SCOPE and RESQUE (Gärtner et al., 2024; Himanen et al., 2024), alongside studies on evaluation practices, academic careers, and research integrity (Quaderi, 2023; Vera, 2017).

On the comparison between Latin America and the rest of the world (SO6), the findings confirm that Latin America is developing its own approach to responsible metrics, closely tied to the open science agenda and concerns about equity and social relevance. Nevertheless, structural asymmetries persist relative to the rest of the world: lower volume and continuity of output, reduced methodological diversity, limited applied research, and narrower thematic coverage. Bridging these gaps will require moving from diagnosis toward empirical and applied research, and more systematically aligning regional agendas with international debates on research assessment reform.

**7. Limitations and future research directions**

This study has several limitations that should be considered when interpreting the findings. The selection of databases with high representation of Spanish and Portuguese literature may favor the visibility of Ibero-American publications, particularly those from Spain, while underrepresenting contributions from other regions and languages. Future research could broaden coverage to additional sources and languages, and explore in greater depth the evaluation frameworks developed in the region, with particular attention to contextualized responsible metrics with potential for transfer to regional and international settings.

**References**

Abramo, G. (2024). The forced battle between peer-review and scientometric research assessment: Why the CoARA initiative is unsound. *Research Evaluation,* rvae021. https://doi.org/10.1093/reseval/rvae021

Abramo, G., D'Angelo, C. A., & Caprasecca, A. (2009). Allocative efficiency in public research funding: Can bibliometrics help? *Research Policy, 38*(1), 206–215. https://doi.org/10.1016/j.respol.2008.11.001

Aguado-López, E., & Becerril-García, A. (2021). Performatividad en la ciencia mexicana: El dispositivo de evaluación del SNI. *Revista Mexicana de Ciencias Políticas y Sociales, 66*(243), 19–53.

Aguillo, I. F. (2021). A response to Delgado-López-Cózar et al.: Impact factor: Many statements and few results. *Profesional de la Información, 30*(3). https://doi.org/10.3145/epi.2021.may.15

Aksnes, D. W., Langfeldt, L., & Wouters, P. (2019). Citations, citation indicators, and research quality: An overview of basic concepts and theories. *SAGE Open, 9*(1), 2158244019829575. https://doi.org/10.1177/2158244019829575

Alberts, B. (2013). Impact factor distortions. *Science, 340*(6134), 787. https://doi.org/10.1126/science.1240319

Aranda, D. (2022). DORA, la exploradora de la evaluación científica: "Season 2". *COMeIN: Revista de los Estudios de Ciencias de la Información y de la Comunicación*, (126), 6.

Arellano-Rojas, P., Calisto-Breiding, C., & Peña-Pallauta, P. (2022). Assessment of scientific research: Improvement of science policies in Latin America. *Revista Española de Documentación Científica, 45*(3). https://doi.org/10.3989/redc.2022.3.1879

Arévalo-Guízar, G., & Rueda-Beltrán, M. (2016). Las revistas académicas: Entre la evaluación y el cumplimiento de su función social. *RELIEVE: Revista Electrónica de Investigación y Evaluación Educativa, 22*(1), 1–16.

Arias, A. H. (2021). Declaración DORA: Naturaleza y objetivos. *Compendium, 24*(47). https://www.redalyc.org/articulo.oa?id=88069714001

Atrio, J. (2018). ¿Cómo perciben los investigadores del CONICET al sistema institucional de evaluación de la ciencia y la tecnología? *Revista Iberoamericana de Ciencia, Tecnología y Sociedad – CTS, 13*(37), 189–229.

Barcelona Declaration on Open Research Information. (2024). *Barcelona declaration on open research information*. https://barcelona-declaration.org/

Bastian, M., Heymann, S., & Jacomy, M. (2009). Gephi: An open source software for exploring and manipulating networks. *Proceedings of the International AAAI Conference on Web and Social Media, 3*(1), 361–362.

Batthyány, K., Vommaro, P., & Rovelli, L. (2023). *Iniciativas y regulaciones multinivel para la ciencia abierta: Infraestructuras abiertas y sistemas de evaluación en Iberoamérica* (Documento de trabajo No. 91). Fundación Carolina.

Becerril-García, A. (2019). *AmeliCA vs Plan S: Mismo objetivo, dos estrategias distintas para lograr el acceso abierto*. AmeliCA. http://www.amelica.org/index.php/2019/01/10/amelica-vs-plan-s-mismo-objetivo-dos-estrategias-distintas-para-lograr-el-acceso-abierto/

Beigel, F., & Salatino, M. (2015). Circuitos segmentados de consagración académica: Las revistas de ciencias sociales y humanas en la Argentina. *Información, Cultura y Sociedad*, (32), 7–31.

Beigel, M. F. (2022). El proyecto de ciencia abierta en un mundo desigual. *Relaciones Internacionales*, (50), 163–181. https://doi.org/10.15366/relacionesinternacionales2022.50.008

Belmar, R. H., & Atenas, J. (2024). Challenges in specialization for research data management in an ethical context. *Acta Bioethica, 30*(2), 319–322. https://doi.org/10.4067/S1726-569X2024000200319

Bhakta, S., Xie, J., Peña, F., & Gillespie, S. H. (2025). Data transparency and reproducibility in health research: Bridging the gap for early-career researchers. *Frontiers in Antibiotics, 4*, 1562002. https://doi.org/10.3389/frabi.2025.1562002

Blondel, V. D., Guillaume, J. L., Lambiotte, R., & Lefebvre, E. (2008). Fast unfolding of communities in large networks. *Journal of Statistical Mechanics: Theory and Experiment*, 2008(10), P10008. https://doi.org/10.1088/1742-5468/2008/10/P10008

Budapest Open Access Initiative. (2022). *BOAI20: Recommendations for the future of open access*. https://www.budapestopenaccessinitiative.org/boai20/

Bordignon, F., Chaignon, L., & Egret, D. (2023). Promoting narrative CVs to improve research evaluation? A review of opinion pieces and experiments. *Research Evaluation, 32*(2), 313–320. https://doi.org/10.1093/reseval/rvad013

Bornmann, L., & Marewski, J. N. (2024). Opium in science and society: Numbers and other quantifications. *Scientometrics, 129*(9), 5313–5346. https://doi.org/10.1007/s11192-024-05104-1

Bourdieu, P. (2008). *Homo academicus*. Siglo XXI Editores.

Brandes, U. (2001). A faster algorithm for betweenness centrality. *Journal of Mathematical Sociology*, 25(2), 163-177.

Butler, L. (2003). Explaining Australia's increased share of ISI publications—The effects of a funding formula based on publication counts. *Research Policy, 32*(1), 143–155. https://doi.org/10.1016/S0048-7333(02)00007-0

Cabezas Clavijo, Á. (2024). What is a narrative CV? 10 tips for its efficient writing. *Infonomy, 2*(2), 15.

Calisto-Breiding, C., Peña-Pallauta, P., & Arellano-Rojas, P. (2021). Transformando la evaluación científica en las políticas de ciencia, tecnología e innovación (CTI) de América Latina y el Caribe: Un estudio desde la altmetría. *Información, Cultura y Sociedad*, (45), 75–94.

Chu, H. (2015). Research methods in library and information science: A content analysis. *Library & Information Science Research, 37*(1), 36–41. https://doi.org/10.1016/j.lisr.2014.09.003

CILAC. (2018). *Declaración de Panamá sobre ciencia abierta*. https://forocilac.org/declaracion-de-panama-sobre-ciencia-abierta/

CLACSO-FOLEC. (2022). *Una nueva evaluación académica y científica para una ciencia con relevancia social en América Latina y el Caribe: Declaración aprobada en la XXVII Asamblea General de CLACSO*. CLACSO.

CoARA. (2022). *Agreement on reforming research assessment*. https://coara.eu/agreement/the-agreement-full-text/

Coccia, M. (2005). A scientometric model for the assessment of scientific research performance within public institutes. *Scientometrics, 65*(3), 307–321. https://doi.org/10.1007/s11192-005-0276-1

CLACSO. (2015). *Declaración de la Asamblea General de CLACSO sobre la evaluación universitaria y científica*. XXV Asamblea General Ordinaria de CLACSO, Medellín, Colombia. https://www.clacso.org/wp-content/uploads/2019/04/6.-Declaracion-MEDELLIN.pdf

Curry, S., de Rijcke, S., Hatch, A., Pillay, D., van der Weijden, I., & Wilsdon, J. (2020). *The changing role of funders in responsible research assessment: Progress, obstacles & the way ahead* (RoRI Working Paper No. 3). https://doi.org/10.6084/m9.figshare.13227914

da Silva, M. R., Rezende, L. V. R., de Campos Ribeiro, G. M., Drumond, L. B. B., da Silva, F. C. C., & Riascos, S. A. C. (2025). Transformations in academic staff evaluation: A study on CoARA in the context of Brazilian universities. *Biblios*, (Esp.), e003. https://doi.org/10.5195/biblios.2025.1254

da Silva, M. R., Silva Santos Rocha, E., Paula Meneses Alves, A., & do Rosário Trindade, D. (2024). Altmetrics and ethics in scientific evaluation: Guidelines, challenges and recommendations for responsible practice. *Biblios*, (87), e002. https://doi.org/10.5195/biblios.2024.1142

Daraio, C., & Vaccari, A. (2022). How should evaluation be? Is a good evaluation of research also just? Towards the implementation of good evaluation. *Scientometrics, 127*(12), 7127–7146. https://doi.org/10.1007/s11192-022-04329-2

de Herde, V., Björnmalm, M., & Susi, T. (2021). Game over: Empower early career researchers to improve research quality. *Insights: The UKSG Journal, 34*, 1–6. https://doi.org/10.1629/UKSG.548

de Rijcke, S., Wouters, P. F., Rushforth, A. D., Franssen, T. P., & Hammarfelt, B. (2016). Evaluation practices and effects of indicator use—A literature review. *Research Evaluation, 25*(2), 161-169. https://doi.org/10.1093/reseval/rvv038

Delgado López-Cózar, E., & Martín Martín, A. (2019). El factor de impacto de las revistas científicas sigue siendo ese número que devora la ciencia española: ¿Hasta cuándo? *Anuario ThinkEPI, 13*(1), 23.

Delgado López-Cózar, E., Rafols, I., & Abadal, E. (2021). Letter: A call for a radical change in research evaluation in Spain. *Profesional de la Información, 30*(3), e300303. https://doi.org/10.3145/epi.2021.may.03

Dirnagl, U., & Pariente, N. (2024). Promoting research quality. *PLOS Biology, 22*(2), e3002554. https://doi.org/10.1371/journal.pbio.3002554

DORA. (2012). *San Francisco declaration on research assessment*. https://sfdora.org/read/

Federation of Finnish Learned Societies, The Committee for Public Information, Finnish Association for Scholarly Publishing, Universities Norway, & European Network for Research Evaluation in the Social Sciences and the Humanities. (2019). *Helsinki initiative on multilingualism in scholarly communication*. https://doi.org/10.6084/m9.figshare.7887059.v1

Gärtner, A., Leising, D., & Schönbrodt, F. D. (2024). Towards responsible research assessment: How to reward research quality. *PLOS Biology, 22*(2), e3002553. https://doi.org/10.1371/journal.pbio.3002553

Hatch, A., & Curry, S. (2020). Changing how we evaluate research is difficult, but not impossible. *eLife, 9*, e58654. https://doi.org/10.7554/eLife.58654

Hicks, D., Wouters, P., Waltman, L., de Rijcke, S., & Rafols, I. (2015). Bibliometrics: The Leiden Manifesto for research metrics. *Nature, 520*(7548), 429–431. https://doi.org/10.1038/520429a

Himanen, L., Conte, E., Gauffriau, M., Strøm, T., Wolf, B., & Gadd, E. (2024). The SCOPE framework: Implementing ideals of responsible research assessment. *F1000Research, 12*, Article 140810. https://doi.org/10.12688/f1000research.140810.2

Jacobs, H. (2024). DORA on steroids. *EMBO Reports, 25*(3), 932–933. https://doi.org/10.1038/s44319-024-00068-y

Jiménez-Contreras, E., de Moya Anegón, F., & López-Cózar, E. D. (2003). The evolution of research activity in Spain: The impact of the National Commission for the Evaluation of Research Activity (CNEAI). *Research Policy, 32*(1), 123–142. https://doi.org/10.1016/S0048-7333(02)00008-2

Johnes, J. (2018). University rankings: What do they really show? *Scientometrics, 115*(1), 585–606. https://doi.org/10.1007/s11192-018-2666-1

Kowaltowski, A. J., Silber, A. M., & Oliveira, M. F. (2021). Responsible science assessment: Downplaying indexes, boosting quality. *Anais da Academia Brasileira de Ciências, 93*, e20191513. https://doi.org/10.1590/0001-3765202120191513

Kulikowski, K., Przytuła, S., & Sułkowski, Ł. (2023). When publication metrics become the fetish: The research evaluation systems' relationship with academic work engagement and burnout. *Research Evaluation, 32*(1), 4–18. https://doi.org/10.1093/reseval/rvac032

López, W. (2013). Declaración de San Francisco: Una vieja discusión desde el contexto latinoamericano. *Universitas Psychologica, 12*(2).

Ma, L. (2022). Metrics and epistemic injustice. *Journal of Documentation, 78*(7), 392–404. https://doi.org/10.1108/JD-12-2021-0240

Melero, R., Uribe Tirado, A., & Armengou, C. (2023). Open scientific knowledge, dissemination and access to academic production: Ideas for a discussion. *Hipertext.net*, (27), 1–4.

Millano-González, V. (2021). Acceso abierto y sistemas de evaluación de la ciencia: ¿Calidad invisibilizada vs. cantidad comercializada? *Revista Técnica de la Facultad de Ingeniería, Universidad del Zulia, 44*(1). https://www.redalyc.org/articulo.oa?id=605772532001

Moher, D., Naudet, F., Cristea, I. A., Miedema, F., Ioannidis, J. P. A., & Goodman, S. N. (2018). Assessing scientists for hiring, promotion, and tenure. *PLOS Biology, 16*(3), e2004089. https://doi.org/10.1371/journal.pbio.2004089

Moher, D., Bouter, L., Kleinert, S., Glasziou, P., Sham, M. H., & Barbour, V. (2020). The Hong Kong principles for assessing researchers: Fostering research integrity. *PLOS Biology, 18*(7), e3000737. https://doi.org/10.1371/journal.pbio.3000737

Morales Perdomo, T. (2022). Editorial. *Tendencia Editorial UR*, (Extra 31), 3.

Orduña-Malea, E., & Bautista-Puig, N. (2024). Research assessment under debate: Disentangling the interest around the DORA declaration on Twitter. *Scientometrics, 129*(1), 537–559. https://doi.org/10.1007/s11192-023-04872-6

Page, M. J., McKenzie, J. E., Bossuyt, P. M., Boutron, I., Hoffmann, T. C., Mulrow, C. D., et al. (2021). The PRISMA 2020 statement: An updated guideline for reporting systematic reviews. *BMJ, 372*, n71. https://doi.org/10.1136/bmj.n71

Quaderi, N. (2023). The misuse of bibliometrics in research assessment infrastructures. *Information Services & Use, 44*(1), 23–26. https://doi.org/10.3233/ISU-230191

Ràfols, I. (2019). S&T indicators in the wild: Contextualization and participation for responsible metrics. *Research Evaluation, 28*(1), 7–22. https://doi.org/10.1093/reseval/rvy030

Rafols, I., & Molas-Gallart, J. (2022). How to reform research evaluation in Spain: Institutional accreditation as a response to the European Agreement on Research Assessment. *Profesional de la Información, 31*(6), e310603. https://doi.org/10.3145/epi.2022.nov.03

Robinson-Garcia, N., Costas, R., Nane, G. F., & van Leeuwen, T. N. (2023). Valuation regimes in academia: Researchers' attitudes towards their diversity of activities and academic performance. *Research Evaluation, 32*(2), 496–514. https://doi.org/10.1093/reseval/rvac049

Rovelli, L. (2022). Hacia una nueva agenda de políticas evaluativas: Aportes desde la iniciativa FOLEC-CLACSO. *Temas de Indicadores de Ciencia y Tecnología 2022*, 21–36. Red Iberoamericana de Indicadores de Ciencia y Tecnología (RICYT). https://dialnet.unirioja.es/servlet/articulo?codigo=8825119

Rushforth, A. (2025). Beyond impact factors? Lessons from the Dutch attempt to transform academic research assessment. *Research Evaluation, 34*, rvaf035. https://doi.org/10.1093/reseval/rvaf035

Salatino, M., & Macedo, A. (2022). El acceso abierto como instrumento para la transformación de la evaluación académica. *Revista Tramas y Redes*, (3), 349–358.

Salatino, M., & Ruiz, O. L. (2021). El fetichismo de la indexación: Una crítica latinoamericana a los regímenes de evaluación de la ciencia mundial. *Revista Iberoamericana de Ciencia, Tecnología y Sociedad – CTS, 16*(46), 73–100.

Saura, G., & Bolívar, A. (2019). Sujeto académico neoliberal: Cuantificado, digitalizado y bibliometrificado. *REICE. Revista Iberoamericana sobre Calidad, Eficacia y Cambio en Educación, 17*(4), 9–26.

Shu, F., Liu, S., & Larivière, V. (2022). China's research evaluation reform: What are the consequences for global science? *Minerva, 60*(3), 329–347. https://doi.org/10.1007/s11024-022-09468-7

Thamhain, H. J. (2014). Assessing the effectiveness of quantitative and qualitative methods for R&D project proposal evaluations. *Engineering Management Journal, 26*(3), 3–12. https://doi.org/10.1080/10429247.2014.11432015

Torres-Salinas, D., Orduña-Malea, E., Delgado-Vázquez, Á., Gorraiz, J., & Arroyo-Machado, W. (2024). Foundations of narrative bibliometrics. *Journal of Informetrics, 18*(3), 101546. https://doi.org/10.1016/j.joi.2024.101546

Torres-Salinas, D., Robinson-García, N., & Jiménez-Contreras, E. (2023). The bibliometric journey towards technological and social change: A review of current challenges and issues. *Profesional de la Información, 32*(2), e320228. https://doi.org/10.3145/epi.2023.mar.28

UNESCO. (2021). *UNESCO recommendation on open science*. UNESCO. https://unesdoc.unesco.org/ark:/48223/pf0000379949

Uribe, A., Vélez, G., & Pallares, C. (2023). Producción científica en Colombia relacionada con ciencia abierta, métricas de nueva generación y métricas responsables en el contexto de Publindex y SCIENTI: Algunas características y perspectivas para apoyar una política nacional. *Revista Científica, 48*(3), 2.

Vasen, F., & Vilchis, I. L. (2017). Sistemas nacionales de clasificación de revistas científicas en América Latina: Tendencias recientes e implicaciones para la evaluación académica en ciencias sociales. *Revista Mexicana de Ciencias Políticas y Sociales, 62*(231), 199–228.

Vélez Cuartas, G. (2022). El imperativo de evaluar la investigación científica y humanística de manera responsable. *Tendencia Editorial UR*, (Extra 31), 4–7.

Vélez Cuartas, G., Suárez Tamayo, M., Jaramillo Guevara, L., & Gutiérrez, G. (2021). Nuevo modelo de métricas responsables para medir el desempeño de revistas científicas en la construcción de comunidad: El caso de Redes. *Redes: Revista Hispana para el Análisis de Redes Sociales, 32*(2), 110–152.

Vélez Cuartas, G., Uribe-Tirado, A., Restrepo-Quintero, D., Ochoa-Gutierrez, J., Pallares, C., Gómez-Molina, H. F., Suárez-Tamayo, M., & Calle, J. (2019). Hacia un modelo de medición de la ciencia desde el Sur Global: Métricas responsables. *Palabra Clave, 8*(2), e068. https://doi.org/10.24215/18539912e068

Vera, H. (2017). El homo academicus y la máquina de sumar: Profesores universitarios y la evaluación cuantitativa del mérito académico. *Perfiles Educativos, 39*(155), 87–106.

Vommaro, P., & Rovelli, L. (2022a). Responsible research evaluation in the university and scientific policy agenda of Latin America and the Caribbean. *Tendencia Editorial UR*, (Extra 31), 8–9.

Vommaro, P., & Rovelli, L. (2022b). Desafíos a la evaluación de la investigación orientada a la movilización del conocimiento en transición hacia la ciencia abierta: Un análisis a partir del caso de los grupos de trabajo del Consejo Latinoamericano de Ciencias Sociales. *Analecta Política, 12*(23), 1–18.

Wilsdon, J. R., Bar-Ilan, J., Frodeman, R., Lex, E., Peters, I., & Wouters, P. (2017). *Next-generation metrics: Responsible metrics and evaluation for open science*. https://eprints.whiterose.ac.uk/113919/

Wilsdon, J., Allen, L., Belfiore, E., Campbell, P., Curry, S., Hill, S., Jones, R., Kain, R., Kerridge, S., Thelwall, M., Tinkler, J., Viney, I., Wouters, P., Hill, J., & Johnson, B. (2015). *The Metric Tide: Report of the Independent Review of the Role of Metrics in Research Assessment and Management*. https://doi.org/10.13140/RG.2.1.4929.1363

Zhao, K., Li, J., & Liang, H. (2026). Reconfiguring power: A field analysis of China's research evaluation reform. *Higher Education, 91,* 1469–1486. https://doi.org/10.1007/s10734-025-01480-6

**AI use statement**

The authors used Claude Sonnet 4.6 (Anthropic, Inc.) to assist with stylistic and grammatical revision of the manuscript, English translation, the generation of Figure 1, and the thematic dimensions and criteria presented in Table 3. All AI-assisted outputs were reviewed and verified by the authors, who take full responsibility for the accuracy and integrity of the work.

**CREDIT Taxonomy**

Daniela Oyarzún-Cristi: Conceptualization, Data curation, Formal analysis, Investigation, Methodology, Visualization, Writing – original draft, Writing – review & editing.

Álvaro Cabezas-Clavijo: Conceptualization, Supervision, Validation, Visualization, Writing – review & editing.

Alicia Moreno-Delgado: Conceptualization, Methodology, Supervision, Validation, Visualization, Writing – review & editing.

**Conflict of interest**

The authors have no competing interests to declare that are relevant to the content of this article.